\documentclass[pdftex,dvipsnames,usenames,aps,superscriptaddress,twocolumn,amsmath,showkeys]{revtex4-2}
\usepackage{graphicx}
\usepackage{color}
\usepackage{amssymb}
\usepackage{amsmath}
\numberwithin{equation}{section}
\usepackage{subcaption}
\usepackage{xcolor}
\usepackage{hyperref}

\def\beq{\begin{equation}}
\def\eeq{\end{equation}}
\def\bea{\begin{eqnarray}}
\def\eea{\end{eqnarray}}
\def\eqref#1{Eq.~(\ref{eq:#1})}
\def\eqlab#1{\label{eq:#1}}

\def\secref#1{Section~\ref{sec:#1}}
\def\seclab#1{\label{sec:#1}}  
\def\figref#1{Fig.~\ref{fig:#1}}
\def\figlab#1{\label{fig:#1}}  
\def\eqref#1{Eq.~(\ref{eq:#1})}

\def\eqlab#1{\label{eq:#1}}

\def\v{{\rm \bf{v}}}
\def\B{{\rm \bf{B}}}
\def\x{{\rm \bf{x}}}
\def\y{{\rm \bf{y}}}

\def\vB{{\rm \bf{v}\!\times\!\bf{B}}}
\def\vvB{{\rm \bf{v}\!\times\!\left(\bf{v}\!\times\!\bf{B}\right)}}

\def\PSF{PKR} 

\newcommand{\Omit}[1]{}

\def\RUG{Kapteyn Astronomical Institute, University of Groningen, P.O. Box 72, 9700 AB Groningen, Netherlands}

\def\VUB{Interuniversity Institute for High-Energy, Vrije Universiteit Brussel, Pleinlaan 2, 1050 Brussels, Belgium}
\def\VUBE{Vrije Universiteit Brussel (VUB), Dienst ELEM, B-1050 Brussels, Belgium}
\def\NIKHEF{Nationaal Instituut voor Kernfysica en Hoge Energie Fysica (NIKHEF), Science Park Amsterdam, Amsterdam, The Netherlands}
\def\IMAPP{Department of Astrophysics/IMAPP, Radboud University Nijmegen, Nijmegen, The Netherlands}

\def\ASTRON{Netherlands Institute for Radio Astronomy (ASTRON), Dwingeloo, The Netherlands}

\def\KIT{KIT, Karlsruhe, Germany}
\def\SoE{Physics Education Department, School of Education, Can Tho University, Campus II, 3/2 Street, Ninh Kieu District, Can Tho City, Viet Nam}

\def\DESY{Deutsches Elektronen-Synchrotron DESY, Platanenallee 6, 15738 Zeuthen, Germany}
\def\ECAP{Erlangen Centre for Astroparticle Physics (ECAP), Friedrich-Alexander-Universit\"{a}t Erlangen-N\"{u}rnberg, 91058 Erlangen, Germany}
\def\Khalifa{Department of Physics, Khalifa University, P.O. Box 127788, Abu Dhabi, United Arab Emirates}

\begin{document}
\title{Aperture correction for Beamforming in radiometric detection of ultra-high energy cosmic rays}
\author{O.~Scholten}  \email[]{o.scholten@rug.nl } \affiliation{\RUG}   \affiliation{\VUB}
\author{T.~N.~G.~Trinh} \email[]{ttngia@ctu.edu.vn}  \affiliation{\SoE}
\author{S.~Buitink} \affiliation{\VUB}
\author{A.~Corstanje} \affiliation{\VUB}
\author{B.M.~Hare} \affiliation{\ASTRON} \affiliation{\RUG}
\author{T.~Huege} \affiliation{\KIT}
\author{B.V.~Jhansi} \affiliation{\Khalifa}
\author{K.~Mulrey} \affiliation{\IMAPP} \affiliation{\NIKHEF}
\author{A.~Nelles} \affiliation{\DESY} \affiliation{\ECAP}
\author{H.~Schoorlemmer} \affiliation{\IMAPP} \affiliation{\NIKHEF}
\author{S.~Thoudam} \affiliation{\Khalifa} 
\author{P.~Turekova} \affiliation{\ASTRON} \affiliation{\RUG}
\author{K.~de~Vries} \affiliation{\VUB} \affiliation{\VUBE}


\begin{abstract}
For high-energy cosmic-ray physics, it is imperative to determine the mass and energy of the cosmic ray that initiated the air shower in the atmosphere. This information can be extracted from the longitudinal profile of the air shower.  In radio-metric observations, this profile is customarily determined through an extensive fitting procedure where calculated radio intensity is fitted to data. Beamforming the measured signals offers a promising alternative to bypass the cumbersome fitting procedure and to determine the longitudinal profile directly.
Finite aperture effects in beamforming hamper the resolution with which this profile can be determined.
We present a comprehensive investigation of the beamforming resolution in radiometric observations of air showers.
There are two, principally different, approaches possible in air-shower beamforming, one where the total beamforming intensity is determined and an alternative where the beamforming trace is cross-correlated with a known response function.
The effects due to a finite aperture (size of antenna array and bandwidth) are large for both approaches.
We argue that it is possible to correct for the aperture corrections using an unfolding procedure. We give an explicit expression for the folding function, the kernel.
Being able to calculate the folding function allows for unfolding the finite aperture effects from the data.
We show that, in a model-to-model comparison, this allows for an accurate reconstruction of the current profile as the shower develops in the atmosphere. We present also an example where we reconstruct the longitudinal current profile of a shower developing under thunderstorm conditions where the atmospheric electric fields greatly alter the orientation of the transverse current in the shower front.

\end{abstract}

\maketitle

\section{Introduction}

The detection of the radio pulse that is emitted by high-energy cosmic-ray or neutrino induced air showers offers a very efficient way to determine the structure of these air showers~\cite{Huege:2016}. Of particular interest for cosmic-ray and neutrino physics is to measure the longitudinal shower profile, the number of particles in the shower as it penetrates in the atmosphere, as this carries information on the energy, the mass of the cosmic ray~\cite{Buitink:2016}, and physics involved in the reactions at the highest energies. The dominant coherent emission of radio waves from such an air shower is induced by the deflection of the charged particles by the geomagnetic field, creating an electric current that is transverse to the shower direction~\cite{Scholten:2008}. This current, driven by the Lorentz force acting on the charged particles in the air shower, will vary with shower depth. The emitted radio signal will thus carry the imprint of the longitudinal profile. Interpreting the measured radio signal, the radio-emission footprint, to reconstruct the longitudinal profile is the central subject of the present work.

The common way for interpreting the radio-emission footprint is through the use of forward modeling~\cite{Buitink:2014}. In a microscopic approach air showers and their radio emission are generated using a Monte Carlo based calculation, where the emission of each particle in the shower is calculated and summed. The available codes for this are CoREAS~\cite{Huege:2013c}, ZHAireS~\cite{Alvarez:2012}, and CORSIKA8~\cite{Alameddine:2023}. The shower that best reproduces the measured footprint is used to obtain the shower profile parameters~\cite{Buitink:2014, AbdulHalim:2024PRL, AbdulHalim:2024PRD}. Another approach is to use the MGMR3D model~\cite{Scholten:2017} where the charge-current cloud in the shower is parametrized and a goodness-of-fit ($\chi^2$) optimization is used to find the parameters of the charge-current cloud that reproduces best the radio-footprint data~\cite{Mitra:2023}. Both procedures are rather indirect where especially the Monte Carlo based methods are very compute intensive. In particular for showers that develop under thunderstorm conditions, where the aim is to learn about the atmospheric electric fields in which the shower develops, the reconstruction of the longitudinal profile requires a complicated fitting procedure~\cite{Trinh:2020,Trinh:2022}.

A more direct procedure, that potentially allows to determine the longitudinal profile model independently is to perform  broad-band beamforming by coherently adding the time-shifted measured pulses in the antennas. The time shifts are chosen to beamform the measured signals to points in the atmosphere that trace the air shower. First steps in this direction were taken by the LOPES collaboration~\cite{Nigl:2008, Apel:2021}. Recently, a more sophisticated procedure  was proposed for air-shower beamforming in~\cite{Schoorlemmer:2021} and independently validated in~\cite{AbdulHalim:2023YS}.
Conceptually, this is very similar to the 3-D beamforming procedure used in lightning imaging~\cite{Shao:2020,Jensen:2021, Scholten:2021-INL} to locate the position of sources that emit broad-band pulses. The main difference between the two is that for lightning imaging special care should be devoted to the polarization of the signal as the antennas are spread over a large area. The polarization of the radio signal is less of an issue for air-shower beamforming, as the arrival direction of the signal is, to a good approximation, the same for all antennas. This 3-D beamforming is often labeled as near-field beamforming to contrast it with beamforming imaging in astronomy where the stars (the sources) are infinitely far and one determines a 2-D position-angle for each star. Just like is the case for astronomical beamforming there are artifacts created in the image due to the fact that there are a finite number of receiving antennas each with a finite aperture. These artifacts show as side-beams and fuzziness in the image. The additional complication in air-shower beamforming, as in any broad-band beamforming of a transient process, is that these artifacts depend on the bandwidth of the receiving antennas.

In this work, we present a calculation of the finite aperture effects on air-shower beamforming. The calculations are based on the generic modeling of an air shower as used in MGMR3D, of which a short review is presented in \secref{MGMR}. The formal calculation of the finite aperture effects in given in \secref{beam}, while in \secref{kern}, we show their effects in semi-realistic examples. In \secref{unfold}, it is shown that to a large extent it is possible to unfold the finite aperture effects from the observations. This deconvolution  be seen as the equivalent of the CLEAN procedure~\cite{Hogbom:1974} used in astronomical imaging, where the difference is that in CLEAN, the folding function is extracted from the data. Particularly interesting is that the procedure reconstructs rather accurate longitudinal profiles for showers that develop under thunderstorm conditions~\cite{Trinh:2020}, where atmospheric electric fields induce very large modifications of the longitudinal current profiles, as well as the generic shower parameters.

\section{Basic formulation using MGMR3D}\seclab{MGMR}

In MGMR3D, a cosmic-ray air shower is modeled as a dynamic three-dimensional charge-current cloud moving towards the ground with the speed of light. The radio emission of this cloud is calculated through the application of Maxwell's equations, see Ref.~\cite{Werner:2012} for an in-depth discussion. This guarantees that all relativistic effects and all coherence effects, linked to the finite extent of the charge-current, are properly accounted for. Much effort has been devoted to the parametrization of this charge current cloud~\cite{Werner:2012, Mitra:2023}, where of particular interest for this work is the transverse four-current density, where the space component is written as
\beq
\rho_x(D_c,h,r_s)= J_x(D_c)\,f(h)\, w(r_s)\;. \eqlab{rho}
\eeq
In \eqref{rho}, $\x$ denotes the direction of the current transverse to the shower axis, given by $\v$, i.e.\ each of the two directions $\x=\vB$ and $\y= \vvB$, where $\B$ is the orientation of the geomagnetic field; $D_c$ denotes the geometric distance of the shower front to the impact point on the ground and is measured along the shower axis, as is indicated in \figref{Kernel-geom}; the shower reaches ground at $t=0$ and $J_x(D_c)$ denotes the total current in the two transverse directions when the shower front is at $D_c$, i.e.,\ at time $t =-D_c/c$; the distance behind the shower front is given by $h$, where $f(h)$ denotes the charged-particle density in the air shower with $f(h<0)=0$; the radial distribution of the current density is given by $w(r_s)$, where $r_s=0$ is on the shower axis and cylindrical symmetry is assumed. The speed of light in vacuum is denoted by $c$, and the index of refraction of air is height-dependent and denoted by $n$ with derivative $n'$. In this work, we assume that the antennas (observers) are in the shower plane, taken perpendicular to the shower axis. Note that for air showers developing under fair-weather conditions, we generally have that the current in the $\vvB$ direction is vanishingly small, and it is thus sufficient to concentrate on $\x=\vB$.

The dominant process for radio emission from air showers is driven by electric currents that are concentrated at the air shower front and are transverse to the shower axis. For fair-weather showers, these currents are induced by the geomagnetic field with an orientation given by the Lorentz force, $\vB$. Under thunderstorm conditions, the strong atmospheric electric fields induce transverse currents that are not aligned with $\vB$ and may change direction with distance along the shower axis. Additionally, there is radio emission due to the fact that there is a net charge excess in the air shower inducing Askaryan radiation~\cite{Askaryan:1962} which is sub-dominant. As discussed in \secref{disc}, we ignore the charge excess radiation in this work. Thus,  only the projection of the electric field on the shower plane contributes to the transverse current.

Limiting to transverse-current emission, the field in an antenna at distance $r_a$ from the core is given by
\begin{equation}
E^x(t_a;r_a)
 =\int \!\! \mathrm{d}h \, \mathrm{d}r_s^2 \, { {\cal S}^x(D_c,h,r_s) \over {\cal D}}
 \; ,
 \eqlab{MGMR_E} \end{equation}
with
\begin{equation}
{\cal D}=(t_a c + D_c)\left| {\mathrm{d}t_a \over \mathrm{d}{D_c}}\right|=n\,\left| R - n \zeta - n'\,R^2\right|
 \;,
\eqlab{denom} \end{equation}
where we have introduced $\zeta={D_c}+h$ as the emission distance of a signal.
Using Eq. (19) from Ref.~\cite{Scholten:2008}, transcribed in the notation used in this work,
\bea
&& E_x(t,d)
 =  \!\! \int \!\! \mathrm{d}h \, \mathrm{d}r_s^2 \, \nonumber \\
 && \left(J_x(D_c) {d\,f(h)\over dh} -
 f(h) \, {J_x(D_c)\over dD_c}\right) \frac{w(r_s)}{\cal D}   \; ,
  \eqlab{MGMR-19}
\eea
we see that the source term can be written as
\bea
&&{\cal S}^x(D_c,h,r_s) =  \frac{\mathrm{d}\,\rho_x(D_c,h,r_s)}{\mathrm{d}h}
\nonumber \\
&=& \left(J_x(D_c) {\mathrm{d}\,f(h)\over \mathrm{d}h} - f(h) \, {\mathrm{d}J_x({D_c})\over \mathrm{d}{D_c}}\right) w(r_s) \;.
\eqlab{Sx} \eea
The arrival time in the antenna $t_a$, shower-front distance ${D_c}$, and emission distance $\zeta$ are related by causality which can be written as
\begin{equation}
0=t_a\,c - D_c + n_\zeta R_\zeta \;,
 \eqlab{G} \end{equation}
with $n_\zeta R_\zeta= n(\zeta) \sqrt{\zeta^2+(\vec{r}_a-\vec{r}_s)^2}$, the optical distance from the source point to the antenna.
The denominator, ${\cal D}$, in \eqref{MGMR_E} results from reducing the Dirac delta function $\delta(L^\mu L_\mu)$ where $L_\mu$ is the optical path~\cite{Werner:2008, Werner:2012}.

For the purpose of this work, it is essential to rewrite the integration in \eqref{MGMR_E} as
\bea
&&E^x(t_a;\vec{r}_a)
 = \int \mathrm{d}\zeta \,\int \mathrm{d}r_s^2 {\cal S}^x(D_c,h,r_s) \frac{1}{n_\zeta R_\zeta} \nonumber \\
 &=& \int\!\! \mathrm{d}D_c\, J_x(D_c) \!\int\!\! \mathrm{d}r_s^2 \frac{{\mathrm{d}\,f(h)\over \mathrm{d}h} w(r_s)}{ n_\zeta^2 (h+{D_c}) -n_\zeta  n_\zeta'\,R^2_\zeta}
 \; ,
 \eqlab{E_x}
\eea
where we used  $\int \mathrm{d}h=\int \left|\mathrm{d}h/\mathrm{d}\zeta\right| \mathrm{d}\zeta$ with $\left|\mathrm{d}h/\mathrm{d}\zeta\right|={\cal D}/(n_\zeta R_\zeta )$, and assume the causality relation \eqref{G}, changed integration variable in the second step, and have omitted the second term in \eqref{Sx} since the first is dominant by four orders of magnitude.

\section{Beamforming}\seclab{beam}

By reverse propagation of the signals in the antennas, \eqref{MGMR_E}, to a focal point at distance ${D_b}$ on the shower axis, we construct the beamforming trace. The trace depends on $t_b$,
\bea
{\cal R}^x(t_b;D_b)&=&\int\!\! \int {E}^x(t';r_a) \frac{{\cal F}\,\delta(G)}{n_b R_b} \, \mathrm{d}t'\, \mathrm{d}^2r_a
\nonumber \\
&=&\int {E}^x(t_a;r_a)\,\frac{{\cal F}}{n_b\,R_b} \, \mathrm{d}^2r_a  \;, \eqlab{R_0}
\eea
where the Dirac delta-function factor, $\delta(G)$, determines the relation between $t_a$ and the the antenna position $\vec{r}_a$,
\begin{equation}
0=G({D_b},t_b;t_a,\vec{r}_a)=t_b\,c - t_a\,c - {D_b} + n_b\,R_b \;,
 \eqlab{G*} \end{equation}
with $n_b\,R_b= n({D_b}) \sqrt{{D_b}^2+\vec{r}_a^2}$.
In principle, an arbitrary function ${\cal F}({D_b},t_b,t_a,r_a)$ can be added inside the integral but for ease of writing we put ${\cal F}=1$.
$t_b=0$ corresponds to the time when the shower front passes distance ${D_b}$ while $t_a=0$ corresponds to the time when the shower front reaches ground.

In actual calculations, ${\cal F}$ is written as a sum of delta-functions, changing the integral over antenna positions to a sum over discrete antenna locations. In principle, a weighting factor could be included to give more weight to antennas with a large signal-over-background ratio. See also the remarks at the end of \secref{disc}.

In the examples presented here, we have limited ourselves to beamforming on the shower axis. Apart from simplifying the expressions, this yields the most sensitive measure of the longitudinal shower profile we are interested in. Off-axis beamforming (not reported on in this work) yields traces that are considerably different, which can be used to determine the position of the shower axis, as shown in~\cite{Schoorlemmer:2024}.

Substituting \eqref{E_x} in \eqref{R_0} we obtain the central equation of the work,
\beq
{\cal R}^x(t_b;D_b) = \int J_x(D_c)\, {\cal K} (t_b;D_b,D_c)\, \mathrm{d}{D_c} \;, \eqlab{RK1}
\eeq
with the kernel
\bea
&&{\cal K}(t_b;D_b,D_c) = \int  \frac{w(r_s)}{n_bR_b} {\mathrm{d}\,f(h)\over \mathrm{d}h} \times \nonumber \\
 && \frac{1}{ n_\zeta^2 (h+{D_c}) -n_\zeta n_\zeta'\,R^2_\zeta} \,\mathrm{d}^2r_a \,\mathrm{d}^2r_s\;, \eqlab{K1}
\eea
where the use of the causality relation, \eqref{G*}, is understood.

\begin{figure}[h]
\centering
\begin{subfigure}{0.23\textwidth}
   \includegraphics[width=0.99\textwidth,bb=0 0cm 15.5cm 17.0cm,clip]{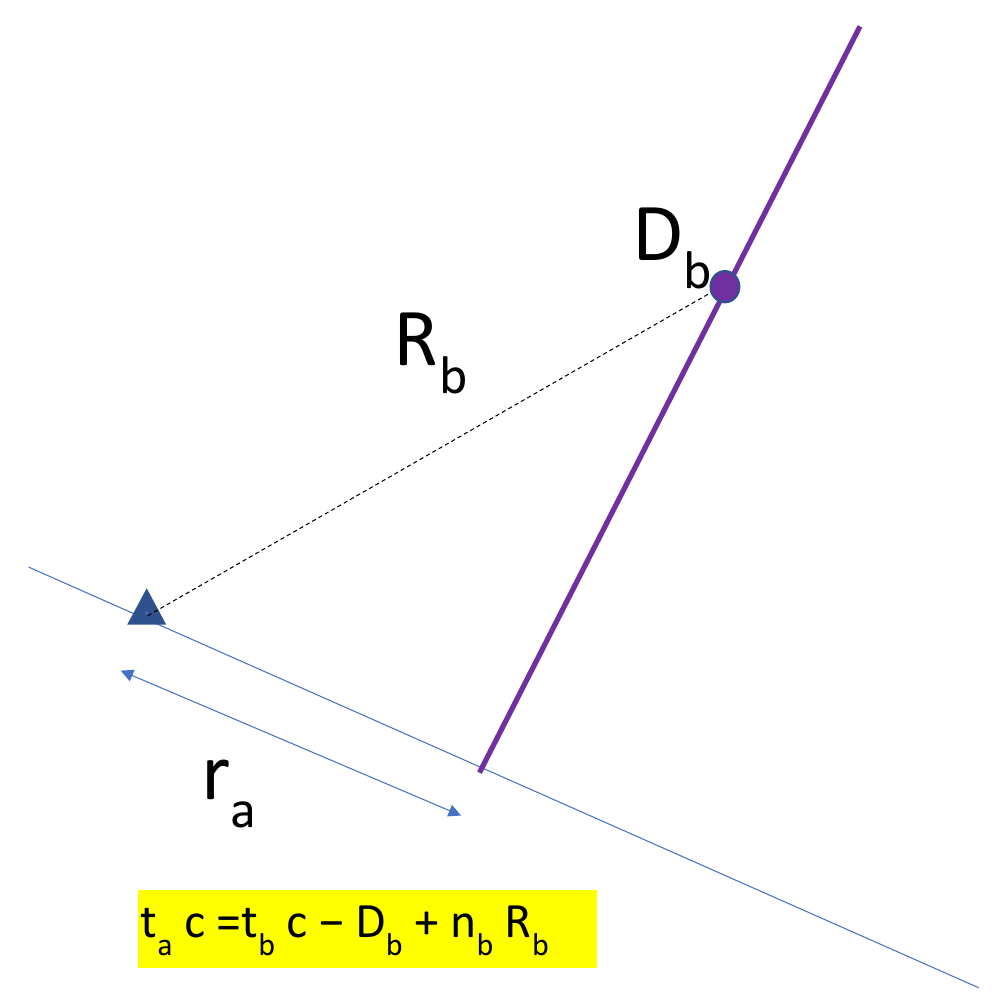}
   \caption{Geometry for Beaming.}
   \figlab{Beam-geom}
\end{subfigure}
\begin{subfigure}{0.23\textwidth}
   \includegraphics[width=0.99\textwidth,bb=0 0cm 15.5cm 17.0cm,clip]{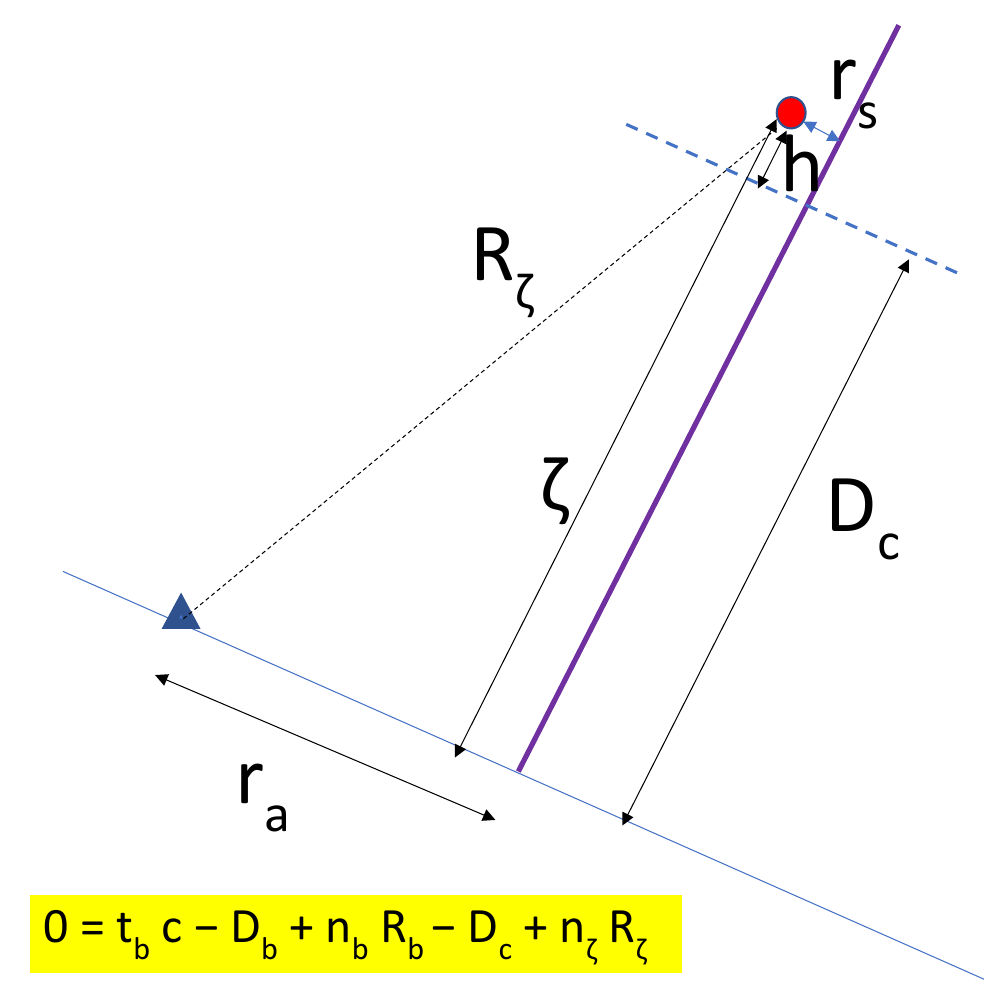}
   \caption{Geometry for kernel.}
   \figlab{Kernel-geom}
\end{subfigure}
\caption{The geometry used for calculating the beamforming trace ${\cal R}^x(t_b;D_b)$ and the kernel ${\cal K}(t_b;D_b,D_c)$. Note that not all indicated parameters are independent since they have to obey the causality relations \eqref{G*} and \eqref{G}, as marked in yellow in the figure.
}
\figlab{geometry}
\end{figure}

The variables used in the integrations are schematically indicated in \figref{geometry}.
To evaluate the Kernel, we have to find the distance $h$ relative to the shower front located at position $D_c$. For this, we use the focal point location parameters $D_b$, $n_b$, and $R_b$ and time $t_b$ to obtain the electric field evaluation time $t_a$ from \eqref{G*}. Having obtained $t_a$, the distance $h$ relative to the shower front, positioned at $D_c$, is found through the causality condition presented in \eqref{G}.

The kernel depends explicitly on the antenna positions as expressed by $\int \mathrm{d}^2r_a$, which may be replaced by a sum over antenna position as noted earlier. The kernel also depends on the shape of the charge cloud, through the functions $f(h)$ and $w(r_s)$ as it moves to ground. As shown in Ref.~\cite{Mitra:2023}, these functions are generic for a large variety of showers. These functions will probably not apply to very inclined showers, where also the assumption of cylindrical symmetry is broken. Cylindrical symmetry is assumed for calculational simplicity, but the approach can easily be generalized. In future work, it needs to be investigated to what extent these assumptions hamper the applicability of the unfolding procedure as discussed in \secref{unfold}. First indications are that these assumptions do not seriously affect the results since in the case of the thunderstorm examples the structure of $f(h)$ used in the model simulation of the `data' differs from that used in the calculation of the kernel used to reconstruct the current profiles.

In principle, the function $w(r_s)$ could, after a re-ordering of the integration order in \eqref{RK1} and \eqref{K1}, be determined following a similar unfolding procedure as discussed in \secref{unfold}. First results (not discussed in this work) indicate that it will be difficult to reach a sufficient accuracy as the emission very close to the shower axis tends to dominate strongly. This can be taken as an argument that thus the detailed form of $w(r_s)$ is not essential. This will be the subject of a future work.

\section{Numerical calculation Kernel}\seclab{kern}

To get some insight in the structure of the kernel, we investigate it through numerical calculations. We show the effects of the limited range of antennas as well as the effects of frequency filtering. In all our calculations, we have taken the shower at a rather arbitrarily chosen angle of 43$^\circ$ from the zenith. The precise value is not important for any aspect of the discussions presented in this work and neither the strength and relative angle of the magnetic field, as all fields and currents are expressed in arbitrary units.

At a fixed beamforming distance $D_b$, the structure of the kernel depends on the layout of the antennas through the explicit integral over antenna position $\vec{r}_a$ as given in \eqref{K1}. \figref{Kernel} shows the structure of the kernel for the case where $D_b=5.5$~km and the antennas cover a distance range of  250~m or 500~m from the shower shower axis in the shower plane (the shower core), i.e\ taking ${\cal F}(\vec{r_a})$ equal to a sum of delta functions in $|\vec{r_a}|$ while keeping the azimuthal integration. A very sharp peak is shown for $t_b\approx 0$ at $D_c\approx D_b$ that extends over a range of about 2~km.
For this study, the antennas are placed in continuous concentric rings at distances of 10~m. Except for small values of $D_b<1$~km, the details of the layout are not important as long as the antennas are spread evenly covering the same area. The structure of the kernel is calculated using a time-step of 0.1~m=0.03~ns in $t_b$.

\begin{figure}[h]
\centering
\begin{subfigure}{0.5\textwidth}
   \includegraphics[width=0.9\textwidth,bb=0 1.8cm 14.5cm 12.8cm,clip]{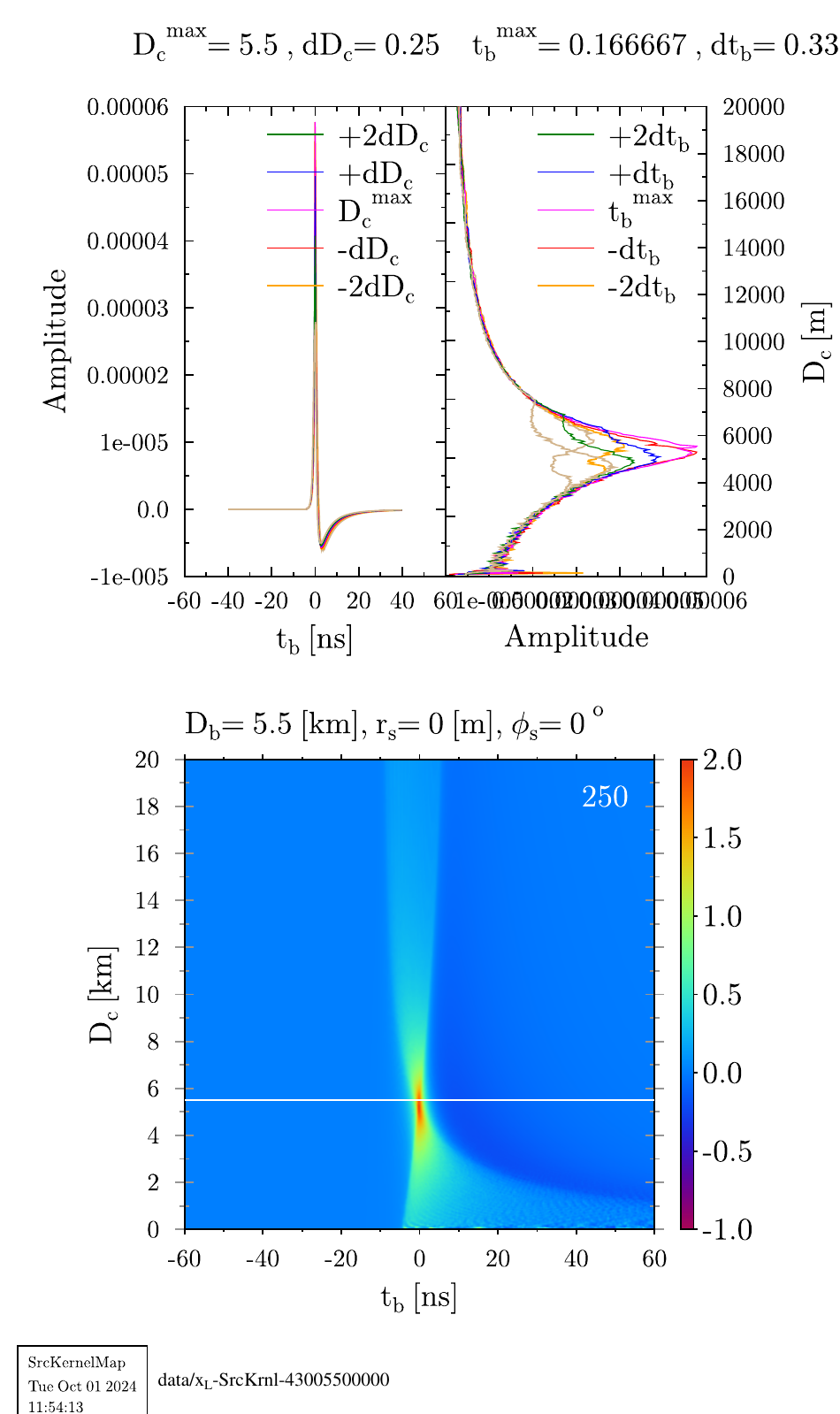}
   \caption{Antenna-range of 250~m.}
   \figlab{Kernel-250}
\end{subfigure}
\begin{subfigure}{0.5\textwidth}
   \includegraphics[width=0.9\textwidth,bb=0 1.8cm 14.5cm 12.8cm,clip]{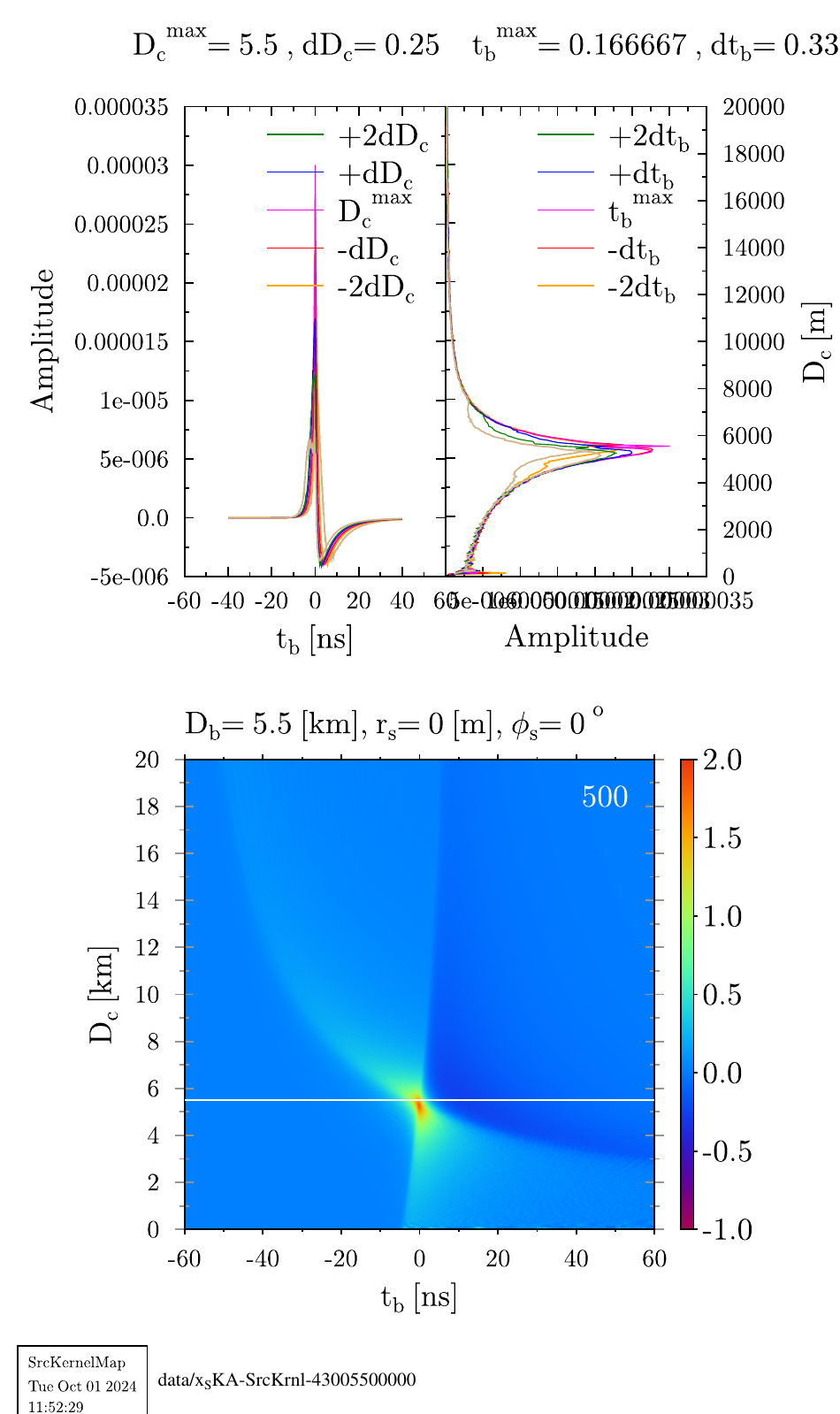}
   \caption{Antenna-range of 500~m.}
   \figlab{Kernel-500}
\end{subfigure}
\caption{The Kernel ${\cal K}(t_b;D_b,D_c)$, see \eqref{K1}, as calculated for the two different ranges of the maximal antenna distance to the shower core in the shower plane. The beaming distance is $D_b=5.5$~km for both panes where $D_b=D_c$ is indicated by the horizontal white line. The color scale is in arbitrary units.
}
\figlab{Kernel}
\end{figure}

The structure of the kernel is -at first sight- very similar for the two distance ranges of 250~m and 500~m as can be seen from \figref{Kernel}. Both show a rather sharp positive-valued spike at $t_b=0$ and $D_c = D_b= 5.5$~km with large wing-like structures extending to larger and smaller values of $D_c$. This tells immediately that for both cases the beamforming signal at distance $D_b$ is influenced by the current in the shower from a large range of distances, $D_c$, extending well below and above $D_b$.
The spike at $D_c = D_b$ is positive and followed by a negative (dark blue) tail extending to large values of $t_b$ because it is formed by the coherent addition of pulses from all antennas that have the same basic structure.
On closer inspection, one sees some important differences. One is that the peak in intensity has become much sharper for the 500~m case. The other is that for values of $D_c \neq D_b$ the `wings' show differently in the two cases.
For $D_c < D_b$ a negative-valued wing extends to larger positive values of $t_b$ for a range of 500~m than is the case for 250~m.
Thus, for antennas covering a smaller range, a current at lower altitudes gives rise to a kernel where the zero-crossing occurs for smaller $t_b$.
This can be understood by the fact that the main positive part of the pulse emitted by a current at low altitudes, arrives relatively later in distant antennas than the pulse emitted from $D_c = D_b$. All these positive contributions from the more distant antennas will thus extend the positive part of the beamforming amplitude to larger $t_b$.
Conversely, for $D_c > D_b$ the emission arrives earlier in the distant antennas than the pulse emitted at $D_c = D_b$, thus giving rise to the much more pronounced wing extending to negative values of $t_b$, as seen by comparing \figref{Kernel-500} with \figref{Kernel-250}.

\begin{figure}[h]
\centering
\begin{subfigure}{0.23\textwidth}
   \includegraphics[width=0.95\textwidth,bb=1cm 2.5cm 13cm 13.7cm,clip]{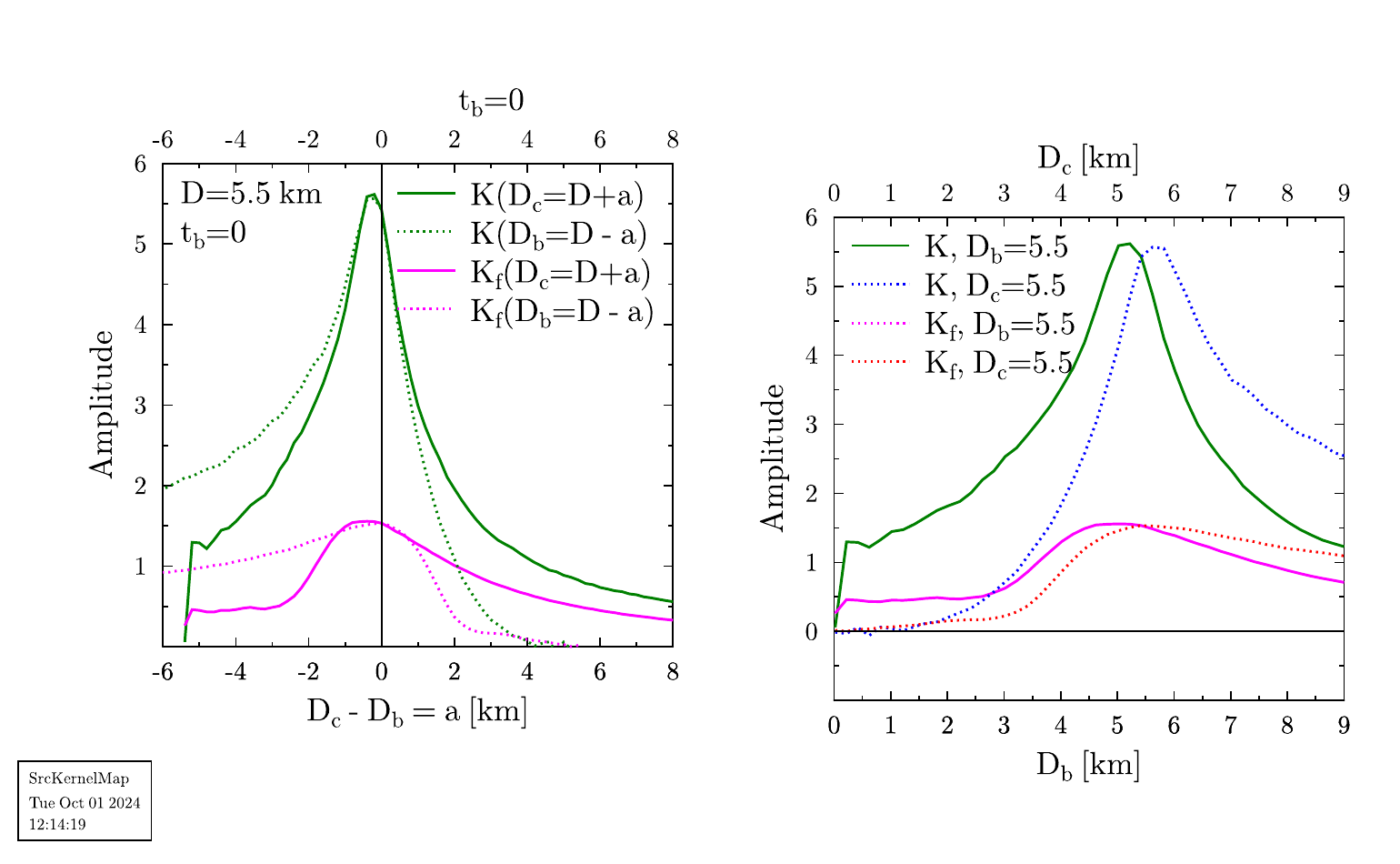}
   \caption{Antenna-range of 250~m.}
   \figlab{KernelFixD-250}
\end{subfigure}
\begin{subfigure}{0.23\textwidth}
   \includegraphics[width=0.95\textwidth,bb=1cm 2.5cm 13cm 13.7cm,clip]{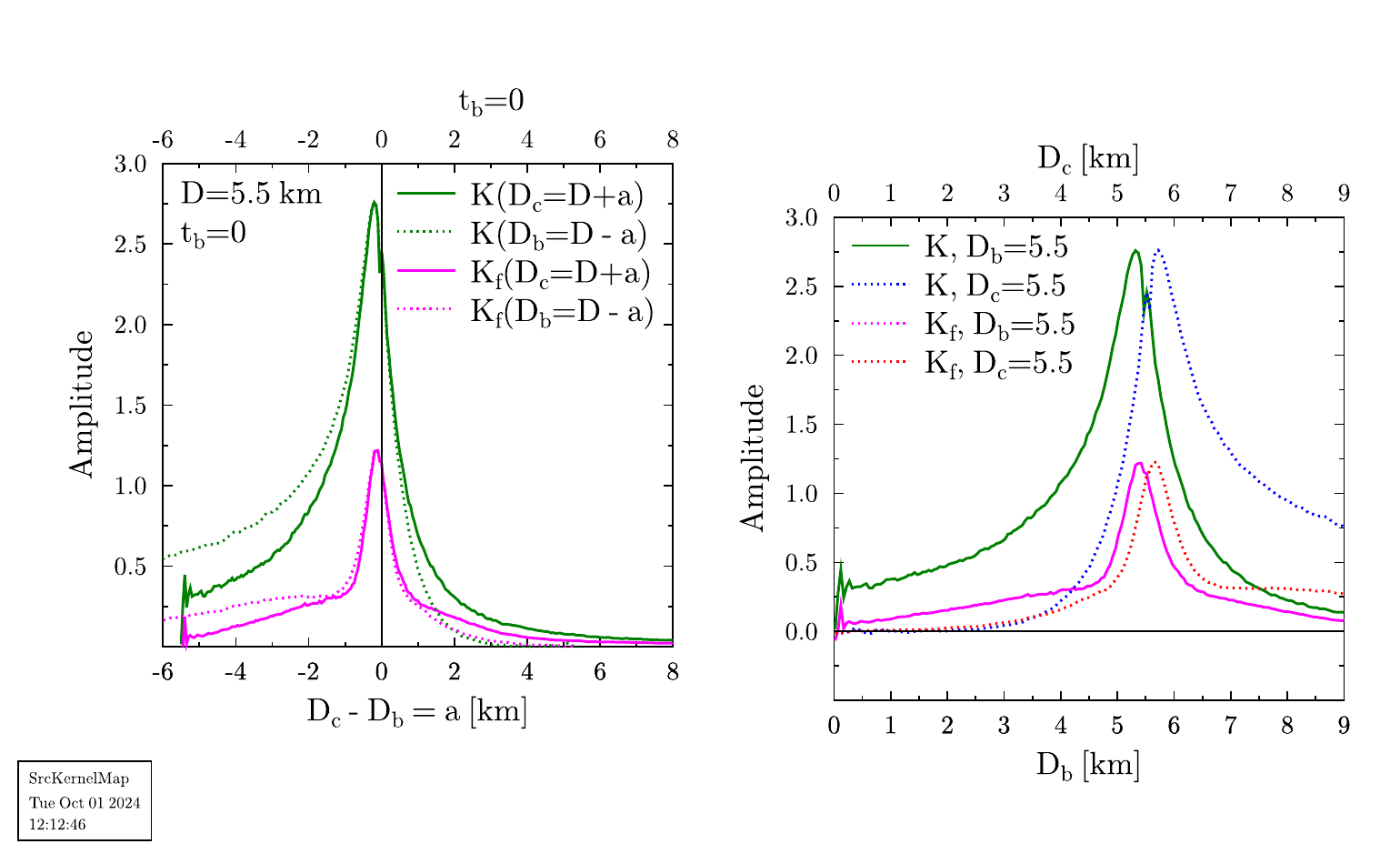}
   \caption{Antenna-range of 500~m.}
   \figlab{KernelFixD-500}
\end{subfigure}
\caption{The value of the kernel ${\cal K}(t_b;D_b,D_c)$ (in arbitrary units) at $t_b=0$ as function of $D_c=D+a$, keeping $D_b=D=5.5$~km fixed (solid curves) and also as function of $D_b=D-a$, keeping $D_c=D=5.5$~km fixed (dotted curves) for two different antenna layouts. The curves in green show the unfiltered kernels, while for the magenta curves in the left (right) panel a 30-80~MHz (50 -- 300~MHz) block filter is applied.
}
\figlab{KernelFixD}
\end{figure}

A more quantitative impression of the effects of the antenna range on the spreading width of the kernel is obtained from \figref{KernelFixD} where we show the dependence of the kernel ${\cal K}(t_b;D_b,D_c)$ at $t_b=0$ around the point $D_c=D_b=D=5.5$~km by varying $D_c-D_b=a$. The green curves show the results for the unfiltered kernel, ${\cal K}(t_b=0;D_b=D-a,D_c=D)$ and ${\cal K}(t_b=0;D_b=D,D_c=D+a)$. The magenta curves, labeled as ${\cal K}_f$ show the filtered kernel using a 30-80~MHz block filter for the antenna range of 250~m (left panel) and 50--300~MHz for 500~m (right)and will be discussed in a following paragraph.
This figure shows clearly that when the antenna array covers a larger range, \figref{KernelFixD-500}, the distribution is much more sharply peaked near $D_c=D_b$ than for a smaller antenna range,  \figref{KernelFixD-250}, precisely as one would expect.

The distinct offset of the peak from $D_b=D_c$ is due to the fact that we have taken $t_b=0$ in \figref{KernelFixD}.
When beaming at a distance $D_b=D=5.5$~km, one `sees' the maximum of the current which is some distance behind the shower front. The shower front thus has reached a slightly lower altitude $D_c < D_b$ (solid curve) due to the finite pancake thickness.
Conversely, when the front of the shower just reached a distance of $D_c=D=5.5$~km, the current at time $t=-D\,c$ ($t_b=0$) peaks at slightly larger altitudes and thus `seen' when $D_b > D_c$ (dotted curve). For the quantitative relation between this shift and the pancake thickness, one should include relativistic beaming effects.
For slightly larger values for $t_b$ the peak is positioned at $D_b=D_c$ (not shown).
Since the integrand for the calculation of the kernel is very non-smooth, due to the sharpness of the shower front near the core, a very fine integration grid was used to obtain \figref{KernelPSF}. With a more coarse integration spurious structures may show that do not affect the final result when frequency filters are applied.

The magenta curves labeled by $K_f$ in \figref{KernelFixD} display the kernel when semi-realistic frequency filters are applied spanning a frequency range one may encounter in realistic scenarios. Reminiscent of LOFAR is the 30 -- 80~MHz block-filter~\cite{Haarlem:2013} applied to the case where antennas up to a range of 250~m, case \figref{KernelFixD-250}, are included.
Reminiscent of an SKA-like scenario is a block-filter of 50 -- 300~MHz~\cite{SKA:2014} for the case of \figref{KernelFixD-500}.
It is seen by comparing the green and magenta curves in \figref{KernelFixD-500} that for the larger antenna range of 500~m the application of a 50 -- 300~MHz block-filter does not greatly affect the peaked structure of the kernel. However, for the more limited antenna range of 250~m, applying the 30 -- 80~MHz block-filter results in a considerable broadening of the peak in the kernel. The main reason for this difference is the effects of the antenna range as can also be deduced from \figref{Kernel}, where the larger antenna range results in a very broad (in $t_b$) beaming trace for $D_b \neq D_c$ as compared to $D_b \approx D_c$ and thus a very fast drop in the value of the filtered kernel.

\begin{figure}[h]
\centering  
   \includegraphics[width=0.3\textwidth,bb=1.3cm 2.5cm 12.4cm 14.0cm,clip]{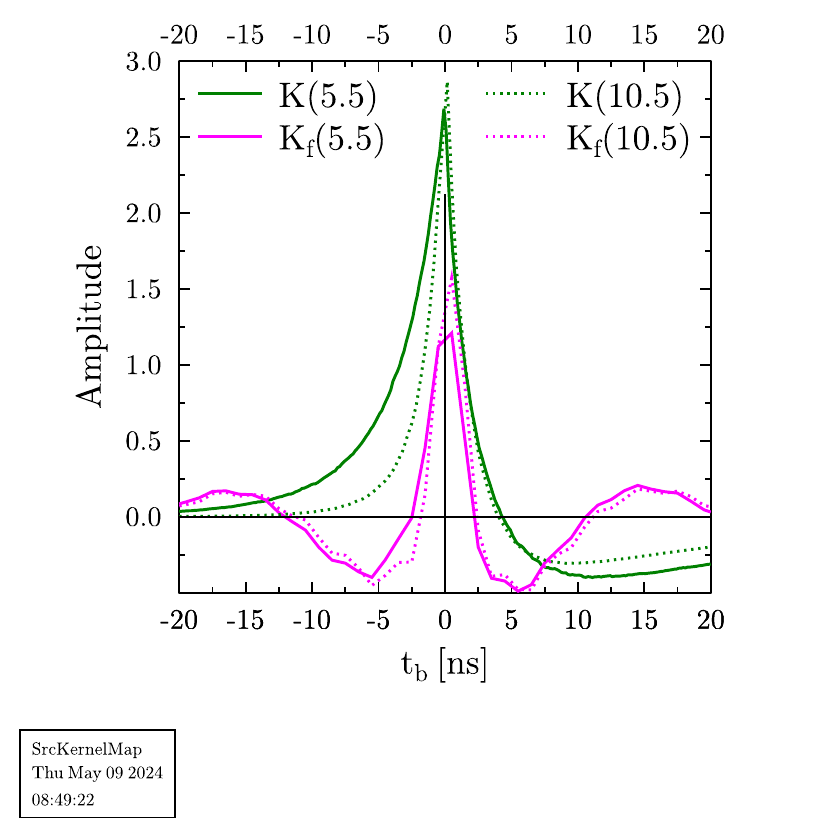}
\caption{The value of the kernel at $D_c=D_b=5.5$~km (solid curves) and at $10.5$~km (dotted curves) as function of  $t_b=0$. The curves labeled by $K_f$ (magenta) display the structure of the kernel when a block frequency filter of 50 -- 300~MHz is applied.
}
\figlab{KernelPSF}
\end{figure}

It can be seen already from \figref{Kernel} that the kernel as a function of $t_b$ exhibits a sharp peak at $t_b\approx 0$ for the case that $D_b\approx D_c$ which gradually becomes less pronounced when the difference between $D_b$ and $D_c$ increases. As is discussed in following sections, this feature will allow to improve the sensitivity of the beaming signal to the shower profile. To this end we show in \figref{KernelPSF} the structure of the peak in the kernel at $D_b = D_c$ for two different distances of 5.5~km (same distance as in the previous figures) and 10.5~km, where we have limited ourselves to the case where antenna range is 500~m from the core. The unfiltered kernel (at a resolution of 0.3~ns) shows a clear peak at $t_b=0$ that is sharper for higher altitudes. The peak is sufficiently sharp that when applying a 50 -- 300 MHz block-filter the resulting time trace resembles closely a sinc-function (not shown) structure, the filtered response to a delta-function, very similar for the two altitudes. For future reference, we name the filtered peak response calculated for the highest altitude the Pulse Kernel Response (\PSF), 
\beq
\PSF(t_b)=N \,{\cal K}(t_b;D_b=D,D_c=D) \;, \eqlab{PFS}
\eeq
where $D$ equals the largest distance considered in the calculation and where $N$ is a normalisation constant such that, $\int |\PSF(t_b)|^2 dt_b=1$.
As shown in \figref{KernelPSF}, the structure of ${\cal K}(t_b;D_b=D,D_c=D)$ does not vary much with $D$, certainly after applying a frequency filter, and the precise value of $D$ is not very relevant. For distances smaller than $D\approx 1$~km, the trace ${\cal K}(t_b;D_b=D,D_c=D)$ starts to differ significantly from the dependence shown in \figref{KernelPSF}, where we have not investigated in detail why this is the case. For this reason the approach is simplified by selecting a single time trace, using the same for all heights.

\section{Extraction of current profile}\seclab{unfold}

In this section we discuss a few different approaches to extract the longitudinal current profile from the beamed antenna traces using the Kernel as discussed in the previous sections. The effectiveness of these approaches will be shown by using MGMR to generate signal traces for all antennas that are subsequently used in beamforming. For the different approaches the extracted longitudinal profile is compared to the one that was used to generate the antenna signals. For fair-weather events, where the current profile has a very simple structure, various extraction methods yield comparable results.
A more severe test case is formed by an air shower where the current profile is strongly influenced by atmospheric electric fields since the direction of the transverse currents may vary greatly with altitude and may even be opposite to that for the fair weather case.

The first step in all approaches is to condense the beamforming trace ${\cal R}^x(t_b;D_b)$, calculated by beamforming the E-field traces of all antennas, into an amplitude. There are two basically different approaches possible.
In the first approach, we extract the beamforming amplitude, ${\cal B}^x(D_b)$, as the square root of the beamforming power,
\beq
{\cal B}^x(D_b) = \sqrt{\sum_{t_b} |{\cal R}^x(t_b;D_b)|^2 } \;. \eqlab{BA}
\eeq
This approach is rather robust since it is not sensitive to an offset in $t_b$.
In the second approach, we use the same beamforming trace ${\cal R}^x(t_b;D_b)$ and cross-correlate it with a pre-defined function, $\PSF(t_b)$, that enhances the importance of the part near $t_b=0$ and may be defined as in \eqref{PFS}. An alternative might be to take a filtered $\delta$-function at $t_b=0$. This yields the \PSF-correlated beaming amplitude
\beq
{\cal P}^x(D_b) = \sum_{t_b} {\cal R}^x(t_b;D_b) \, \PSF(t_b) \;. \eqlab{Px}
\eeq
In this approach, care should be taken to correct for any offsets in determining $t_b$ which could be done by finding the offset in $t_b$ that maximizes ${\cal P}^x$ (not done in this work). By enhancing the part of the beamforming trace at $t_b=0$, the contribution near $D_b=D_c$ is increased. An additional advantage is that the information on the sign of ${\cal R}^x(t_b;D_b)$ is kept, which is lost in the procedure of \eqref{BA}. An other advantage of the \PSF-correlation approach is that it suppresses the effects of noise in the data (not considered in the present work), while for ${\cal B}^x(D_b)$ one would need an explicit correction.
In principle, it is possible to keep the full beamforming trace ${\cal R}^x(t_b;D_b)$ and not reduce it to an amplitude, but in this exploratory work it is more insightful to explore simplifications.

The second step is the extraction of the longitudinal current profile from the beamforming amplitude. As discussed in the following two sections, this may be achieved following two different approaches, all based on the use of \eqref{RK1}. In one an analytic parametrization is used, while in the other a more agnostic parametrization. Both approaches are discussed extensively in the following two sections.

To perform more detailed tests of the `data', the E-field traces of all antennas, are generated from an MGMR3D simulation for this work. The beamforming is performed separately for the $\x=\vB$ and $\vvB$ polarization directions. For simplicity, it is assumed in the present study that the antennas are arranged in densely-packed concentric rings in the shower plane centered at the shower core at 10~m separation. The data-traces are time sampled as would be the case in a real measurement.
It is essential that the kernel, ${\cal K} (t_b;D_b,D_c)$ see \eqref{K1}, is calculated for the same antenna layout as used in the 'data', using the same frequency filtering and the same sampling time for $t_b$. The distances $D_b$ and $D_c$ are calculated for a 50~m grid spacing. The kernel depends on the geometry, i.e.\ shower angle and antenna layout, while the simulated E-fields, the 'data', depend also on the shower profile.
We use the fact that the structure of the charge-current cloud, in MGMR3D defined by the functions f(h) and w($r_s$), are rather universal shower parameters. In the discussion in \secref{disc} we will return to this point.

\subsection{Using the beamforming amplitude}\seclab{CurrKern}

To extract the current profile from the beamforming amplitude, determined from the measured electric fields in the antennas, we use a very general parametrization of the current profile $J_x(D_c)$ and optimize the parameters by minimizing
\beq
\chi^2=\sum_{D_b} \left(\frac{{\cal B}_K^x(D_b) - {\cal B}^x(D_b)}{\sigma_B} \right)^2 \;, \eqlab{RMS-K}
\eeq
using a steepest descent method where ${\cal B}^x(D_b)$ is the beamforming amplitude constructed from the data as defined in \eqref{BA}, $\sigma_B$ the estimate of the uncertainty and where
\beq
{\cal B}_K^x(D_b) = \sqrt{\sum_{t_b} |{\cal R}_K^x(t_b;D_b)|^2 }
\eeq
is the modeled beamforming amplitude where the modeled beamforming trace is obtained from a simple folding procedure,
\beq
{\cal R}_K^x(t_b;D_b) = \sum_{D_c} J_x(D_c)\, {\cal K} (t_b;D_b,D_c)\;.
\eeq
Since for the present examples we work with simulated data we have put $\sigma_B=1$. Note that one should contract the current with the kernel before taking the the sum of the squares. Reversing this order would allow to simplify the procedure, however, introduces some approximations that make little difference for fair-weather showers but are severe for air showers under thunderstorm conditions. We will not pursue such an approach in this work.

In an alternative approach, one may minimize the root-mean-square differences between the traces instead of the amplitudes. This will probably improve the sensitivity of the approach, however, we have not pursued such an approach.

We have used a parameterization of the current profile based on the Gaisser-Hillas formula~\cite{Gaisser:1977,Andringa:2011,Aab:2019}, where we have opted for a generalized parametrization in terms of the $R$ and $L$ parameters,
\beq
J^x(X)= \sum_{m=1}^{N_I}  I^x_m \left(1-R^x\, \frac{X^x_m-X}{L^x}\right)^{{R^x}^{-2}} \, e^{\frac{X^x_m-X}{L\,R}} \;,
\eqlab{RLN}
\eeq
where $I^x_m$ is the current strength at penetration depth $X=X^x_m$ and $N_I$ typically equals 1 or 2 for fair weather showers. $N_I=2$ allows for `double bump' showers of the kind shown in Fig. 9 of Ref.~\cite{Mitra:2023}, where the longitudinal shower profile may have more than one clear maximum. As a simplification, mainly to limit the number of parameters, we have taken the values for the $R^x$ and $L^x$ parameters independent of $m$. The parameters may differ for the two polarization directions, where for fair weather showers we take $\x=\vB$ exclusively. The penetration depth and distance to ground are related by the structure of the atmosphere (taken equal to the US standard atmosphere parametrized by Linsley~\cite{Heck:1998}) and the zenith angle of the cosmic ray taken equal to 43$^\circ$, quite arbitrarily.

\begin{figure}[h]
\centering  
   \includegraphics[width=0.40\textwidth,bb=0.0cm 4.3cm 30.5cm 37.5cm,clip]{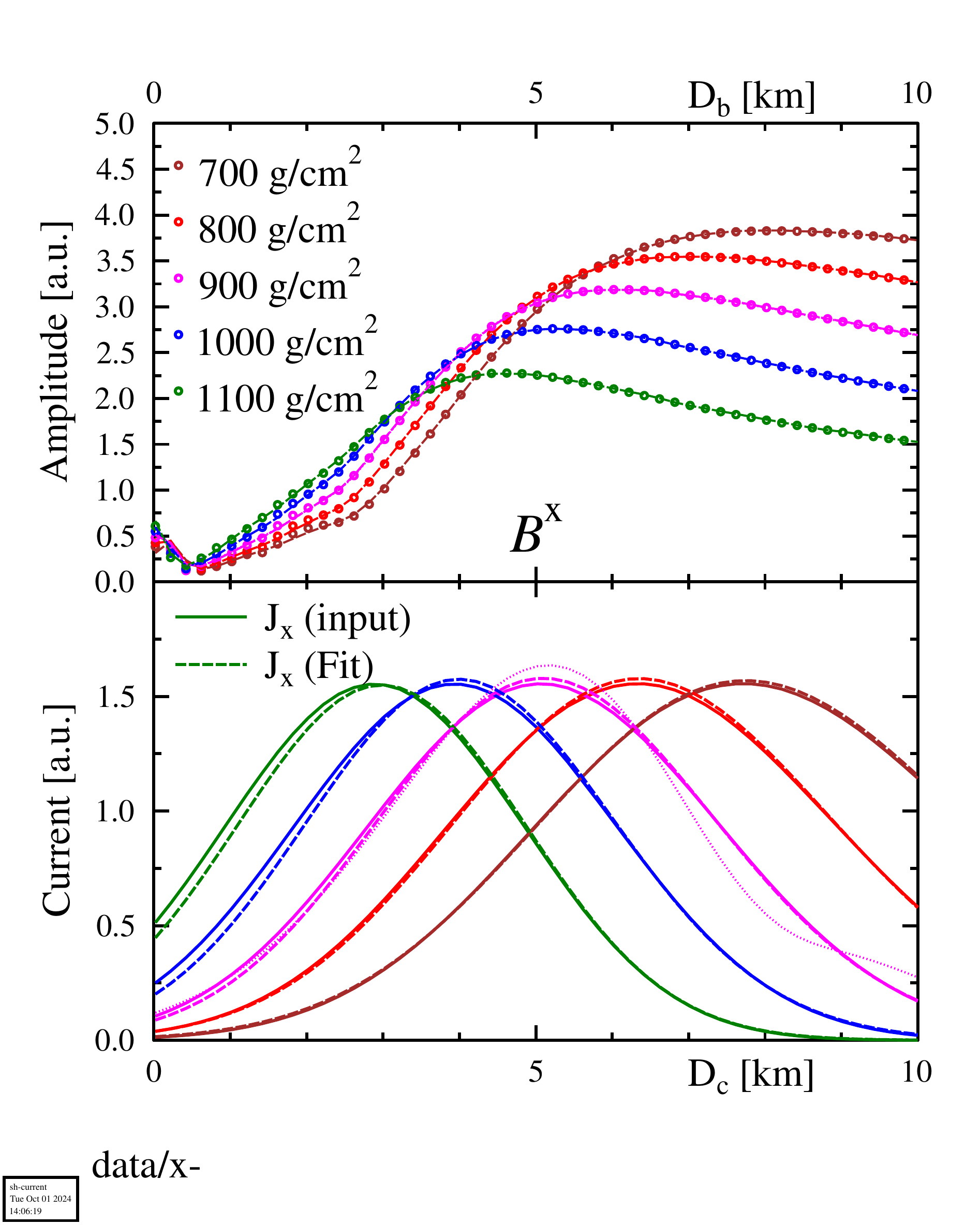}
\caption{The top panel gives the beamforming amplitude ${\cal B}^x(D_b)$ as defined in \eqref{BA} for five showers with different values for $X_{max}$ as indicated in the legend. The bottom panel displays the longitudinal current profile $J_x(D_c)$ for the five different showers (solid curves) as well as the the extracted profile based on using \eqref{RMS-K} (dashed lines) using a single Gaisser-Hillas profile ($N_I=1$ in \eqref{RLN}). The dotted curve shows for the case of $X_{max}=900$~g/cm$^2$ the result of the procedure using \eqref{RLN} with $N_I=3$. Antennas up to a range of 250~m have been included as well as a 30 -- 80~MHz block filter.
}
\figlab{Curr_K_250}
\end{figure}

As an example, the results for the reconstructed currents, using \eqref{RMS-K} with \eqref{RLN} are shown in \figref{Curr_K_250} using the more limited range of 250~m for the antennas and a 30 -- 80~MHz block filter for calculating the fields. The top panel gives the beamforming amplitudes that show very limited structure towards the larger distances for all five values of $X_{max}$, the penetration depth at which the number of charged particles in the shower reached a maximum,  ranging from 700 -- 1100$~g/cm^2$. For comparison, $X_{max}=$ 800$~g/cm^2$ corresponds to $D=6.2$~km and a Cherenkov radius of about 90~m in the antenna plane. The solid lines in the bottom panel show the longitudinal current profiles for the five different cases, while the dashed lines show the reconstructed profiles using $N_I=1$ in \eqref{RLN} thus fitting the four parameters, $I^x_1$, $X^x_1$, $R^x$ and $L^x$ . Due to the fact that the filtered kernel is rather structure-less, as can be seen from \figref{KernelFixD-250}, this number of parameters is close to the limit that can be extracted, even for this rather idealized case where no realistic noise has been added. Increasing the number of parameters to 8 by taking $N_I=3$ in \eqref{RLN} results in clear over fitting, as displayed by the dotted curve in the bottom panel for $X_{max}=900~g/cm^2$. The fit to the beamforming amplitudes, a dotted curve in the top panel of \figref{Curr_K_250} is indistinguishable from the one generated for $N_I=1$ (dashed-dotted).

\subsection{Parametrized current profile using the \PSF}\seclab{CurrPSF}

An alternative approach is to cross-correlate the beamforming trace, obtained from the data, with the pulse-shape function given in \eqref{PFS} yielding the \PSF-correlated beaming amplitude as was defined in \eqref{Px}, ${\cal P}^x(D_b) = \sum_{t_b} {\cal R}^x(t_b;D_b) \,\PSF(t_b) $. Performing the same operation on the kernel, we obtain a weighting function
\beq
{\cal W}(D_b,D_c) = \sum_{t_b} {\cal K} (t_b;D_b,D_c) \, \PSF(t_b) \;. \eqlab{W}
\eeq
Following a similar approach as in the previous section, we obtain the current profile $J^x(D_c)$ by minimizing
\beq
\chi^2={\sum_{D_b} \left|\frac{\left[\sum_{D_c}{\cal W}(D_b,D_c) \, J^x(D_c) \right] - {\cal P}^x(D_b)}{\sigma_P} \right|^2} \;, \eqlab{RMS-PSF}
\eeq
where we have put $\sigma_P=1$ as is more appropriate for a model-to-model comparison. This expression is numerically much simpler than the equivalent expression \eqref{RMS-K} since it is at most quadratic in $J^x(D_c)$.

To solve for $J^x(D_c)$, we have investigated two different parameterizations of the current profile, one based on the Gaisser-Hillas parametrization, \eqref{RLN}. As was done in \secref{CurrKern}, we will use a steepest descent method to solve for the minimal $\chi^2$ value.

A second solution method is using a piece-wise linear (PWL) parametrization, where the current is parametrized by its values on a grid with a linear interpolation for the in-between points. With this parametrization the expression for $\chi^2$, \eqref{RMS-PSF}, reduces to \eqref{RMS-PWL} where the derivatives of the $\chi^2$ are linear in the PWL-parameters. The minimization condition can thus be written as a matrix multiplication that can be solved analytically, see Appendix \ref{sec:PWL}, resulting in
\beq
J_i = A^{-1} B \;, \eqlab{PWL-fit}
\eeq
where $i$ labels the grid points, and the matrices $A$ and $B$ are defined in \eqref{A} and \eqref{B}. The number of parameters for this case is equal to the grid points for defining the current.

\begin{figure}[h]
\centering  
   \includegraphics[width=0.40\textwidth,bb=0.0cm 4.3cm 30.5cm 37.5cm,clip]{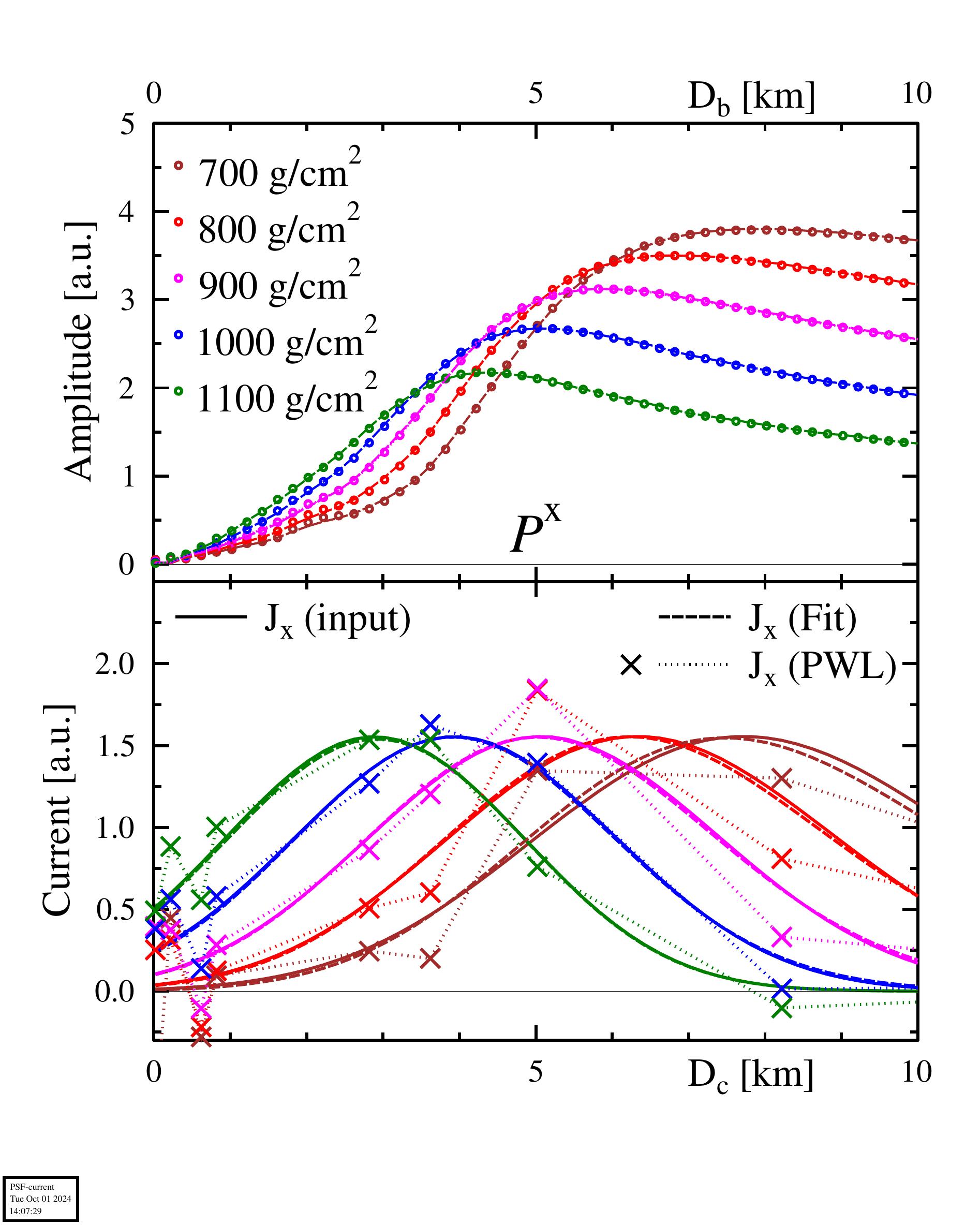}
\caption{The top panel gives the \PSF-correlated beamforming amplitude ${\cal P}^x(D_b)$, as defined in \eqref{Px}, for five showers with different values of $X_{max}$ as indicated in the legend. The bottom panel displays the longitudinal current profile $J_x(D_c)$ for the five different showers (solid curves) as well as the the extracted profile based on using \eqref{RMS-PSF} (dashed lines) using a single Gaisser-Hillas profile. The dotted lines show the results of the PWL calculation, using \eqref{PWL-fit}. Antennas up to a range of 250~m have been included as well as a 30 -- 80~MHz block filter. The ordinate is in arbitrary units for both panels.
}
\figlab{Curr_PSF_250}
\end{figure}

\figref{Curr_PSF_250} shows the results for the extracted current profiles using the \PSF-correlation approach. Using the analytic parametrization for extracting the current profile gives results that closely match the the current profile used for calculating the E-field traces in the different antennas. The quality of the agreement is similar to the the beamforming-amplitude approach, shown in \figref{Curr_K_250}.

The currents can also be extracted using the PWL parametrization. This has the advantage that it contains no assumptions concerning the structure of the current profile, and additionally that the parameters can extracted using the analytic procedure of \eqref{PWL-fit}. The major disadvantage, as is shown clearly in \figref{Curr_PSF_250}, is that it is prone to over-fitting resulting in saw-tooth patterns. In the PWL results shown in \figref{Curr_PSF_250}, we have used a grid that is more dense at smaller distances where the weight functions \eqref{W} have a more pronounced structure. In all cases the fits of the Gaisser-Hillas as well as the PWL fits to the \PSF-correlated beamforming amplitudes, ${\cal P}^x(D_b)$, shown in the top panel of \figref{Curr_PSF_250}, are almost indistinguishable. Care should be taken in choosing the PWL grid points; A too dense grid will give rise to results that show large zigzagging around the correct average. The PWL grid points in \figref{Curr_PSF_250} are denoted by the crosses and have been taken more dense for values of $D_c$ where the kernel shows more structure. The PWL fitting procedure can probably be improved by taking a more sophisticated procedure for selecting the grid and devising some kind of penalty procedure for a strongly fluctuating result. The deviations seen in \figref{Curr_PSF_250} appear to be dependent on $X_{max}$ and it thus might be that a more sophisticated procedure for the grid points, where the variation in the \PSF-correlated beamforming amplitude is taken into account, is worth investigating.

\begin{figure}[h]
\centering  
   \includegraphics[width=0.40\textwidth,bb=0.0cm 4.3cm 30.5cm 37.5cm,clip]{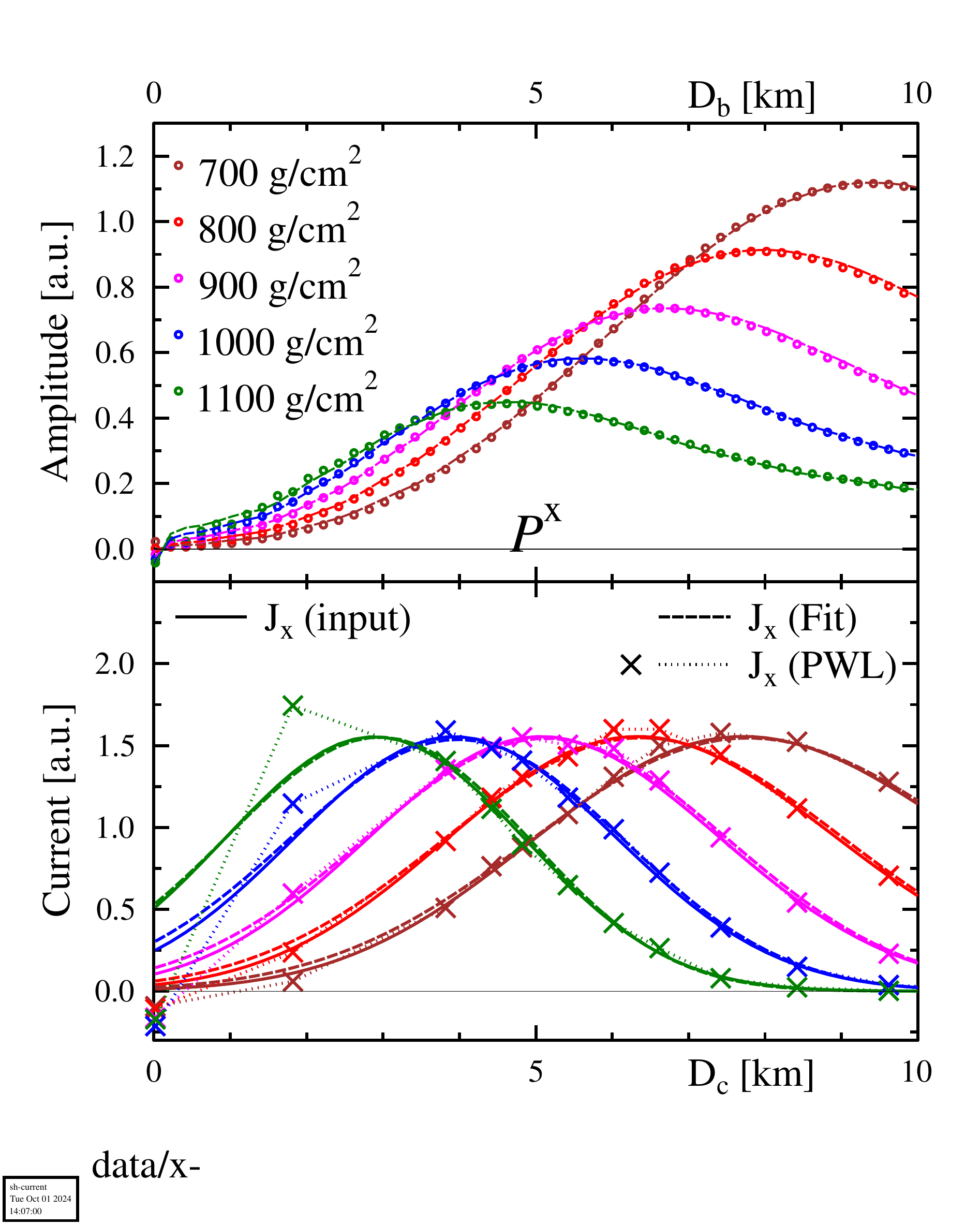}
\caption{Same as \figref{Curr_PSF_250} but for the case where the antenna range is 500~m and a 50 -- 300~MHz block filter is used.
}
\figlab{Curr_PSF_SKA}
\end{figure}

Applying the two extraction procedures to the case where the antennas cover a larger range and a larger bandwidth, shown in \figref{Curr_PSF_SKA}, gives results that are qualitatively very similar to those shown in \figref{Curr_PSF_250} although the results using the PWL-parametrization show considerably less scatter. This is a reflection of the fact that the spreading-width of the kernel for the larger antenna range is considerably smaller than for the small antenna range, as was shown in \figref{Kernel}.

\begin{figure}[h]
\centering  
   \includegraphics[width=0.40\textwidth,bb=0.0cm 4.3cm 30.5cm 37.5cm,clip]{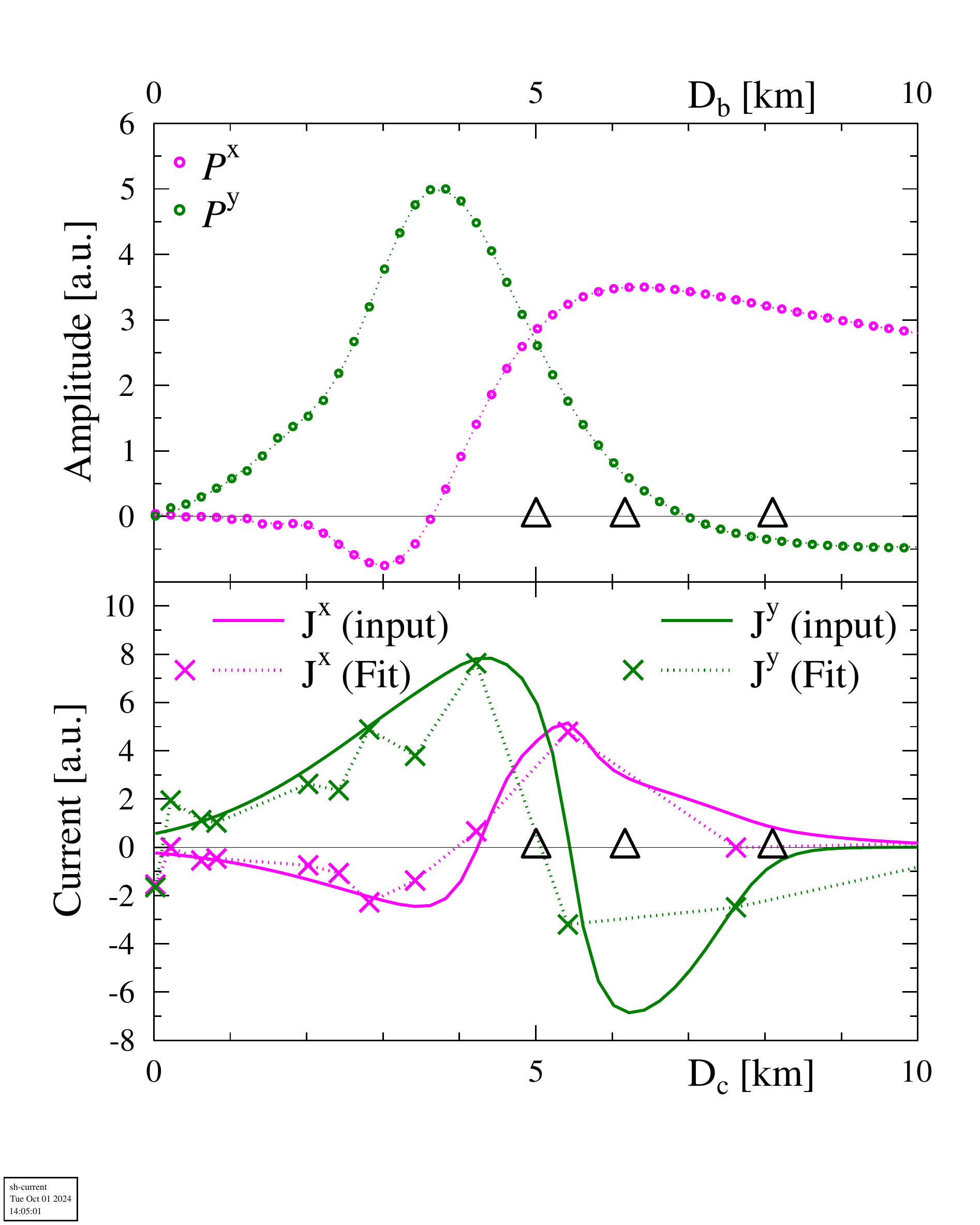}
\caption{The top panel gives the \PSF-correlated beamforming amplitude, \eqref{Px} for $\x=\vB$ and $\y=\vvB$ polarization directions for a cosmic-ray shower developing under influence of strong atmospheric electric fields as discussed in the text. The bottom panel displays the longitudinal profiles for the currents in the two directions. The solid curves show the profiles as used in the simulations while the crosses mark the extracted results obtained by using the PWL parametrization. The triangles mark the boundaries of the E-field layers. Antennas up to a distance of 250~m have been included as well as a 30 -- 80~MHz block filter.
}
\figlab{Curr_Th_250}
\end{figure}

The most stringent test of the current profile extraction is for the case in which the shower develops in an atmosphere with strong electric fields as is the case under thunderstorm conditions. As mentioned in the second paragraph of \secref{MGMR}, we are mainly sensitive to the component of the electric field that is transverse to the shower axis.
To investigate this more complicated shower profile, we have performed a simulation using a typical three-layered structure for the atmospheric electric field as found in Refs~\cite{Trinh:2017, Trinh:2020, Trinh:2022}. Specifically, we have used an atmospheric electric field in the top layer, between distances to the shower core of 8.1 and 6.2~km, with a projection in the shower plane of 103~kV/m at an angle of $\alpha=-77^\circ$ with respect to the $\x=\vB$ direction, a middle layer between 6.2 and 5~km distance, with a projection in the shower plane of 100~kV/m at $\alpha=66^\circ$, and a bottom layer from 5~km to the ground with a field of 108~kV/m at $\alpha=118^\circ$ in the shower plane. For the cosmic ray, we have taken the same geometry as for the other examples in this work with $X_{max}=900$~g/cm$^2$ (corresponding to $D=5$~km). In order to extract the current profiles, we thus have used the same kernels as used in the previous examples.

The \PSF-correlated beamforming amplitudes, as well as the extracted currents, are shown in \figref{Curr_Th_250}. Since the atmospheric electric fields vary greatly in angle in the three different layers the  currents are oriented in rather different directions transverse to the shower axis resulting in non-vanishing beaming traces extracted for the $\vB$ as well as the $\vvB$  polarization directions. Also notable is that ${\cal P}$ is no longer positive definite as was the case for fair-weather showers. This is due to the fact that the current at some distances may be opposite to the direction of the Lorentz force, $\vB$. This is indicating the importance of using ${\cal P}$ as ${\cal B}$, calculated from the absolute square of the beamforming trace, is not sensitive to the sign of the current.

We observed that using \eqref{RLN} to extract the current profile leads to rather unstable results and this method will thus not be explored any further for the case of atmospheric electric fields.
This instability of the fit using the Gaisser-Hillas parametrization for the current profile is due to the fact that the current profile has considerably more structure than a typical fair-weather profile. This structure is difficult to capture in the parametrization of \eqref{RLN} even using $N_I=3$ and the fit gets stuck in a local minimum. A reasonable result can be obtained only by carefully choosing the starting values in the fit and this approach is thus not applicable in general.
We have not explored if there are other analytic parametrizations possible that yield more stable results. The PWL-parametrization offers, in contrast, very reliable results for this case as is seen in \figref{Curr_Th_250}. Even though the extracted current profile shows scatter, it traces the main features well.

\begin{figure}[h]
\centering  
   \includegraphics[width=0.40\textwidth,bb=0.0cm 4.3cm 30.5cm 37.5cm,clip]{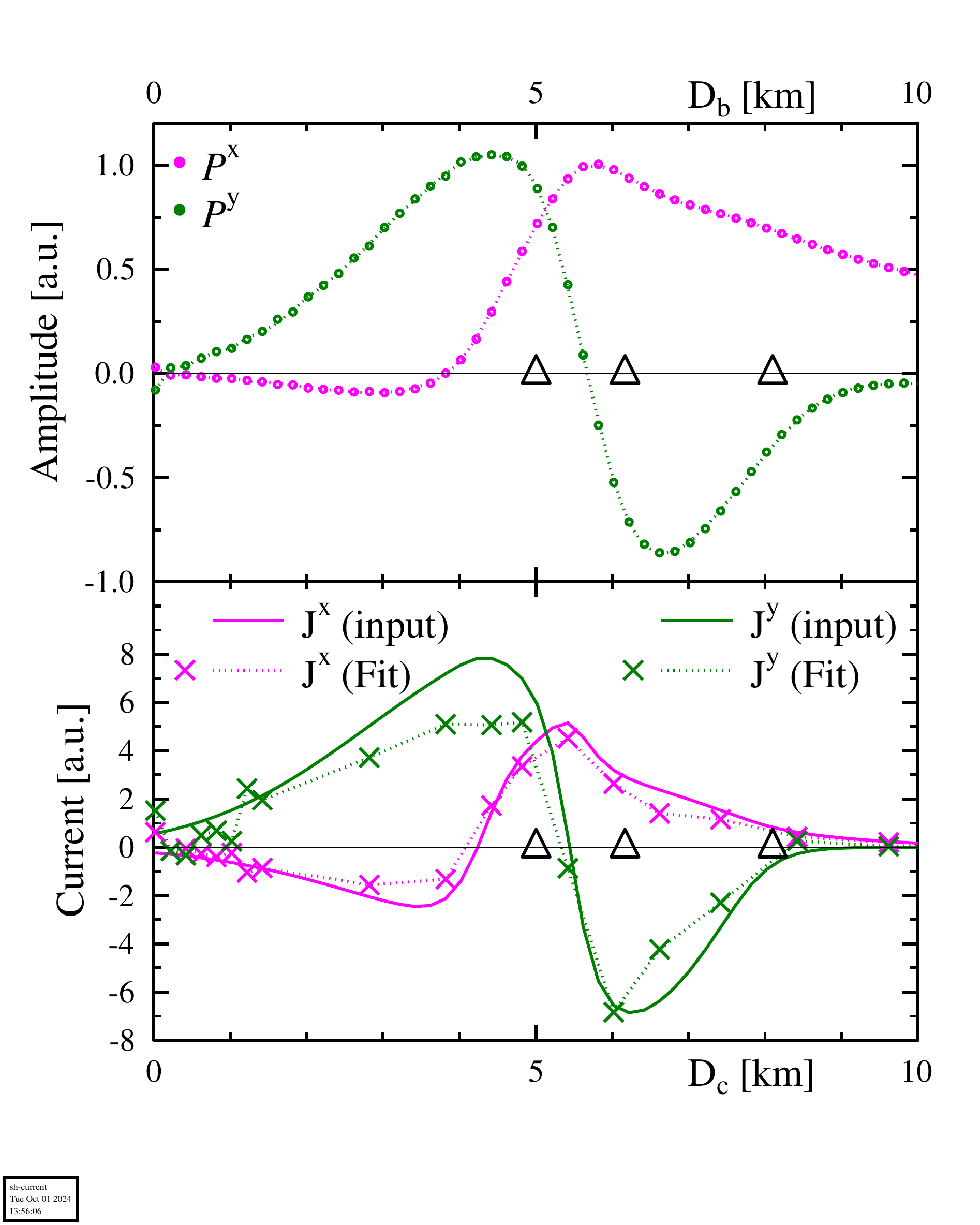}
\caption{Same as \figref{Curr_Th_250} but for the case where the antenna range is 500~m and a 50 -- 300~MHz block filter is used.
}
\figlab{Curr_Th_SKA}
\end{figure}

Using a larger antenna range and a larger bandwidth gives results that are even more accurate as shown in \figref{Curr_Th_SKA}. The current in the simulation (solid curves on the bottom panel) is the same as used in \figref{Curr_Th_250}. With the increased range of antennas and the increased bandwidth, the sensitivity to the current profile at larger distances has greatly improved.

\section{Summary and Discussion}\seclab{disc}

One of the prime motives for radiometric observations of extensive air showers is to be able to extract the longitudinal intensity profile.
Using beamforming, i.e.\ the coherent addition of the signals of all antennas, offers, potentially, a very sensitive method. However, due to the finite range of distances covered combined with the finite band-width of the antennas, the resolution of the procedure is hampered.
We have investigated the effects of these aperture limitations. We have shown that these effects can be modeled using MGMR3D which gives an accurate description of the generic shower properties~\cite{Mitra:2023}. This resulted in the central equation of this paper, \eqref{E_x}. Based on this, we presented different procedures that allow to unfold the effects of the limited aperture and reconstruct the longitudinal shower profiles in a macroscopic model-to-model comparison.

We have investigated two different methods for using beamforming, one where the total beamforming intensity is determined and an alternative where the beamforming trace is cross-correlated with a known response function. It is shown that both procedures allow for an accurate reconstruction of the longitudinal current profile using a generic Gaisser-Hillas parametrization that is known to give an accurate parametrization of the profile for fair-weather showers.

We have also explored a PWL parametrization of the current. This has the advantage that it is agnostic. For the case of fair weather showers, it appears more advantageous to implement in the fit that the current is varying rather smoothly with distance along the shower axis. For showers developing under the influence of (strong) atmospheric electric fields, as are present in thunderstorms, we find in contrast that the agnostic PWL approach gives more satisfactory results. It should be noted that the PWL parametrization is just one of the many possibilities. A different combination of a flexible parametrization combined with a well-chosen cost function will likely yield improvements.

In the present work, we have limited ourselves to transverse-current initiated radio emission thereby excluding charge-excess radiation. From observations it is known that the charge-excess radiation is sub-leading. 
The charge-excess radiation is polarized in the radial direction which interferes with the linearly polarized emission by the transverse current emission. This interference is the reason for the broken cylindrical symmetry in the fluency pattern. In spite of its marked effect on the structure of the radio footprint there are reasons for not considering charge-excess radiation in this work. One reason is, that by summing the $\vB$ or the $\vvB$ polarization components for antennas in a ring around the core, the radial polarization component cancels and thus does not contribute to the analysis. To obtain an effect of charge excess radiation, one thus needs to introduce a particular asymmetric antenna layout, or integrate the radial polarization direction, which we defer to a discussion in a follow-up work where we plan to make a direct link with experiment.
Another reason is that analyzing charge excess radiation requires a different expression for the kernel. Exploratory calculations (not reported here) show that the kernel for charge excess exhibits a much less pronounced peak near $D_c=D_b$ than the kernel for transverse current emission. The reason for this is probably that for charge excess radiation the intensity near the core is strongly suppressed. This results in a less accurate determination of the emitting current. Combined with the fact that charge excess radiation is sub-leading compared to transverse current emission does imply that for real experiments it is doubtful that beamforming the charge excess component will yield useful results.

In interpreting the longitudinal current profile, one should be careful that this will be different from the longitudinal charged-particle profile that is most often considered in air-shower physics. For fair-weather events, they are closely related as the drift velocity, defined as the ratio of the two, is a very smooth function of air density~\cite{Mitra:2023}. Under thunderstorm conditions, the two differ strongly as the current, in magnitude and direction, is dominated by the atmospheric electric fields.

From the onset, we have limited the paper to on-axis beamforming for air showers. The formalism can easily be extended to off-axis beamforming, see the discussion at the end of \secref{unfold}.

In a future work, we plan to apply the proposed procedures to microscopic Monte-Carlo model predictions and real data such as measured at LOFAR.
Attention points will be the investigation of the effects of charge-excess radiation, uncertainty in the core position, and arrival direction on the beamforming results. These will be particularly difficult to correct for with an uneven distribution of antennas over the area.
One attention point for a realistic scenario will thus be the effect of the detailed antenna layout.

In this work, we have shown the importance of including larger distances, where we have used a relatively homogeneous coverage of the area and thus a large number of antennas at large distances. For a non-homogeneous coverage, as one will have in real measurements, there may be an excess antenna density at some radius which will thus dominate the beaming trace leading to worse resolution in $D_c$. It may be thus beneficial to compensate for this by adjusting the function ${\cal F}({D_b},t_b,t_a,r_a)$  introduced in \eqref{R_0}. However, there is a balance as a larger antenna density improves the signal-to-noise ratio, where there exists an extensive discussion in the literature, see for example Refs.~\cite{Briggs:1999, Yatawatta:2014}, for the pros and cons of the different choices. 

\section{Conclusion}\seclab{disc}

Our investigation has demonstrated that it is possible to correct for finite aperture effects when beamforming the radio-frequency pulse emitted by air showers. This greatly improves the resolution with which the longitudinal profile can be extracted from measurements. As an example, one may compare the results shown in \figref{Curr_K_250}. The top panel shows the results from beamforming when no aperture correction is applied. The maximum of the beamforming amplitude for $X_{max}=1100$~g/cm$^2$ lies close to $D=4.5$~km or at about $950$~g/cm$^2$ with a profile shape that does not resemble that of a realistic shower profile. When applying the procedure proposed in this work the extracted $X_{max}$ differs from the input value by only 9~g/cm$^2$ with a very reasonable result for the longitudinal profile function. In spite of the fact that much of the off-sets are systematic in nature the improvement of the procedure when applying aperture corrections is evident.
The results show that beaming might be the ideal approach for air shower measurements at SKA, where a fairly homogeneous antenna coverage over a large area is combined with a large frequency band width.

\begin{acknowledgments}
BMH and PT are supported ERC Grant Agreement No.\ 101041097.
KDdV is supported by the European Unions Horizon 2020 research and innovation program, Grant No.\ 805486.
This work was supported by Deutsche Forschungsgemeinschaft (DFG, German Research Foundation), Projektnummer 531213488.
\end{acknowledgments}

\appendix
\section{Piece-Wise Linear fit}\seclab{PWL}

In a piece-wise linear parametrisation of the current it is written as a sum of partially overlapping triangles where the base of the triangle with the peak at $D_i$ stretches from $D_{i-1}$ till $D_{i+1}$,
\beq
F_i(D)= \left\{ \begin{array}{cl}
\frac{D-D_{i-1}}{D_i-D_{i-1}} &{\rm when }  \;\; D_{i-1} < D < D_i \;,  \\
\frac{D_{i+1}-D}{D_{i+1}-D_{i}} &{\rm when } \;\; D_{i} < D < D_{i+1} \;, \\
0 &{\rm otherwise }  \;.
\end{array}\right.
\eeq
The current is subsequently written as
\beq
J(D)= \sum_{i=1}^N F_i(D) \, J_i \;. \eqlab{J_PWL}
\eeq
In terms of this current the expression for the quality factor becomes after some minor rewriting
\beq
\chi^2_{\rm PWL}
=\int\!\! \mathrm{d}D_b \,\left| \sum_{i=1}^N J_i  {\cal W}_i(D_b) - {\cal P}(D_b) \right|^2
\;, \eqlab{RMS-PWL}
\eeq
with ${\cal W}_i(D_b)=\int_{D_c}\! \mathrm{d}D_c \,{\cal W}(D_b,D_c) \, F_i(D_c)$ .
The minimum is found by setting the derivatives $\partial{\chi^2_{\rm PWL}}/\partial J_j=0$ resulting in
\beq
B_j=\sum_i A_{j,i}\, J_i \;, \eqlab{B=AJ}
\eeq
with
\beq
B_j=\int\!\! \mathrm{d}D_b\,{\cal P}(D_b)\, {\cal W}_j(D_b) \;, \eqlab{B}
\eeq
and
\beq
A_{j,i}=\int\!\! \mathrm{d}D_b\, {\cal W}_j(D_b)\, {\cal W}_i(D_b) \;. \eqlab{A}
\eeq
\eqref{B=AJ} can be solved for $J_i$ by a simple matrix inversion.
For applying this procedure it is not necessary that the values $D_i$ form a regular grid.

\bibliography{} 

\end{document}